\documentclass{PoS}

\title{Quantum corrections to broken $\mathbf{N=8}$ supergravity}

\ShortTitle{Quantum corrections to broken $N=8$ supergravity}

\author{\speaker{Fabio Zwirner}\thanks{Work supported in part by the ERC Advanced Grant \textit{``Electroweak
      Symmetry Breaking, Flavour and Dark Matter: One Solution for
      Three Mysteries''}  (\textit{DaMeSyFla}) and by the Initial
    Training Network  \textit{``Unification in the LHC Era''}  (\textit{UNILHC}).}\\
        Department of Physics and Astronomy, University of Padua,
        Italy \\ Istituto Nazionale di Fisica Nucleare, Sezione di Padova, Italy \\
        E-mail: \email{fabio.zwirner@pd.infn.it}}

\abstract{The discovery of the Higgs boson and the non-discovery (so far) of additional particles at the TeV scale underline our poor understanding of the hierarchy problems in the physics of the fundamental interactions.
Loosely motivated by this consideration, I review some recent results on the classical and quantum stability of Minkowski vacua in spontaneously broken $N=8$ supergravity.}

\FullConference{36th International Conference on High Energy Physics,\\
		July 4-11, 2012\\
		Melbourne, Australia}

\begin{document}

\section{Introduction and motivations}

This talk is based on some recent work \cite{dz} with G.~Dall'Agata. Because of the mixed audience and of the time constraints, I will briefly describe the motivations, the main results and the open problems. Technical details and a more complete list of references can be found in \cite{dz}.

The Higgs boson discovery opening this Conference marked the triumph not only of the LHC machine and of the ATLAS and CMS experiments, but also of the Standard Model (SM) as a renormalizable quantum field theory. With no direct or indirect signals yet of some additional physics at the TeV scale, it is very difficult to understand the status of the SM as an effective theory and the nature of its ultraviolet completion. Data so far have not been very encouraging for the various forms of new physics at the TeV scale, proposed over the years to explain the stability of the Fermi scale of weak interactions with respect to the higher physics scales in the fundamental theory underlying the SM. Part of the theory community is trying to understand how new physics has been hiding from us so far but could be just lurking around the corner. This approach was unsuccessful at LEP and at the Tevatron, but could nevertheless be successful at the LHC with more data at higher energy.  Other theorists have tried to explore, in different ways, the possibility that the new physics scale may be connected with quantum gravity. Including gravity, of course, puts the naturalness problem in a different light, since, for a given ultraviolet cutoff and according to our present  understanding, the scale of the vacuum energy density is far more unnatural than the scale of electroweak symmetry breaking.

On the basis of the above considerations, there may be a point in studying theories that contain gravity, as long as they are constrained enough that well defined questions, with calculable and unambiguous answers, can be posed. String theory is a potential candidate but, when questions about the stability of the vacuum are asked, we have to face the still poor understanding of compactification and supersymmetry breaking in a truly stringy context. With such a state of affairs, a constrained and calculable toy theory, even if non-realistic, could be the playground where to learn some useful lessons. And among the four-dimensional theories including gravity, $N=8$ supergravity stands out as the simplest and most constrained one. 

\section{$N=8$ supergravity and its gaugings}

The first striking property of $N=8$ supergravity (to be considered, here and in the following, in $D=4$ space-time  dimensions) is its unique field content: there is a single multiplet, the gravitational multiplet, containing 128 bosonic and 128 fermionic degrees of freedom. A simple way of constructing the gravitational multiplet is to start from the graviton state of helicity $+2$, and to act on it repeatedly with the 8 anti-commuting supercharges $Q_i$ ($i=1,\ldots,8$).  So doing we obtain:
\begin{eqnarray}
| +2 \rangle : & & 1~graviton \, , 
\nonumber \\
| +3/2 \, , \, i \rangle =  Q_i \, | +2 \rangle : & & 8~gravitinos \, , 
\nonumber \\
| +1 \, , \, [ij] \rangle =  Q_i \, Q_j | +2 \rangle : & & 28~vectors \, , 
\nonumber \\
| +1/2 \, , \, [ijk] \rangle =  Q_i \, Q_j \, Q_k | +2 \rangle : & & 56~fermions \, , 
\nonumber \\
| 0 \, , \, [ijkl] \rangle =  Q_i \, Q_j \, Q_k \, Q_l | +2 \rangle : & & 70~scalars \, , 
\label{multiplet}
\end{eqnarray}
and similarly for the CPT-conjugate states of negative helicity, ending with $| -2 \rangle = \prod_{i=1}^8  Q_i  \, | +2 \rangle $.

In the so-called {\it ungauged} formulation \cite{ungauged}, the theory has just gravitational interactions and a huge duality group, $E_{7(7)}$. This duality generalizes the one already present in the Maxwell theory of electromagnetism: it is an invariance of the system of equations of motion plus Bianchi identities, and exploits the fact that in four dimensions we can have dual electric and magnetic vector fields. The gauge group is trivial, $U(1)^{28}$, with no charged fields. The classical equations of motion admit flat four-dimensional space, with unbroken $N=8$ supersymmetry, as a solution, there is no potential and all the fields are massless. The perturbative expansion about this background displays some remarkable properties (for a recent review and references, see {\it e.g.} \cite{finiteness}): the theory is on-shell finite up to four loops, and it has been conjectured that it could be finite to all orders in perturbation theory.  

In the so-called {\it gauged} versions of $N=8$ supergravity, a subgroup $G$ of $E_{7(7)}$, of dimension $d \le 28$, is made local. As spelled out in \cite{gauged}, the theory is fully determined by the so-called {\it embedding tensor} $\Theta_M{}^\alpha$, which specifies how the local gauge generators $X_M$ are defined in terms of the generators $t_\alpha$ of the duality group:
\begin{equation}
	X_M = \Theta_M{}^\alpha \, t_{\alpha} \, .
\end{equation}
In the above expresssion, the index $M=1,\ldots,56$ runs over all the electric and magnetic vectors in the theory, whilst $\alpha=1,\ldots,133$ runs over the adjoint representation of $E_{7(7)}$. Consistency with supersymmetry and gauge invariance imposes some linear and quadratic constraints on the embedding tensor $\Theta$, whose general solutions, the {\it gaugings}, and the resulting vacua have not been classified yet. The effect of a gauging is to introduce in the theory a deformation parameter, the gauge coupling constant $g$, which appears in the covariant derivatives:
\begin{equation}
\partial_\mu 
\quad \longrightarrow \quad 
D_\mu \equiv \partial_\mu - g \, A_\mu^M \, X_M \, .
\end{equation}
As a result, a scalar potential and mass terms are generated, and the possibility arises of partial or total supersymmetry breaking, with critical points that can have positive, vanishing or negative energy density. Despite the absence of a full classification, a good number of examples are known. An intriguing fact is that, despite decades of efforts, no locally stable classical de-Sitter vacuum has been found so far.  

\section{Some recent results}

In our recent study \cite{dz}, we considered the gaugings leading to classical Minkowski vacua with fully broken ($N=0$) supersymmetry, and examined the one-loop corrections to the theory around these vacua. The arguments that justify the remarkable ultraviolet behaviour of the ungauged theory do not apply to the gauged theory, thus the first question that can be asked is whether the gauged theory is one-loop finite and whether it admits one-loop stable Minkowski or de-Sitter vacua. 

The basic tool to answer such questions is the one-loop effective potential, which in turn is comtrolled by the {\it supertraces} of the generalized mass matrices, evaluated on the classical background. Their definition is
\begin{equation}
{\rm Str} \, {\cal M}^{2k}  \equiv  \sum_a (2J_a +1) \, (-1)^{2J_a} \,  (M^2_a)^k \, , 
\end{equation}
where $k=0,1,2,\ldots$, the index $a$ runs over the different particles in the spectrum, $M^2_a$  and $J_a$ are the corresponding squared-mass eigenvalues and spins. The above equation applies to an arbitrary theory with any number of massive spin-3/2 gravitinos, spin-1 gauge bosons, spin-1/2 Weyl fermions and spin-0 real scalars. Of course, massless particles do not contribute to the supertraces. It is easy to see, starting from the standard textbook expression \cite{cw}, that ${\rm Str} \, {\cal M}^0$, ${\rm Str} \, {\cal M}^2$ and ${\rm Str} \, {\cal M}^4$ control the quartic, quadratic and logarithmic divergences of the one-loop effective potential, respectively. 

The coefficient of the one-loop quartic divergence,  ${\rm Str} \, {\cal M}^0$, is always field-independent, and counts the number of bosonic minus fermionic degrees of freedom, thus it vanishes in all theories with linear representations of supersymmetry. The new general result in \cite{dz}  is the proof that the supertraces of the quadratic and quartic mass matrices also vanish at {\em all} classical four-dimensional Minkowski vacua with spontaneously broken $N=8$ supergravity: 
\begin{equation}
\label{result1}
{\rm Str} \, {\cal M}^2 = {\rm Str} \, {\cal M}^4 =0 \, .
\end{equation}
Skipping all the technicalities, the essential ingredients of the proof are the critical point condition on the classical potential, the vanishing of the classical vacuum energy density and the quadratic constraints on the emebedding tensor. This result makes the one-loop effective potential calculable and finite, for all tachyon-free constant field configurations along the classically flat directions. 

To put the result in context, it should be kept in mind that until 2011 the only known gauging of $N=8$, $D=4$ supergravity leading to classically stable Minkowski vacua with fully broken supersymmetry was the one obtained in 1979 by Cremmer, Scherk and Schwarz (CSS) \cite{CSS}, whose one-loop quantum properties were studied soon after \cite{qcCSS}.  However, a recent paper \cite{di} (see also \cite{more}) introduced a new convenient method for generating $N=8$ gaugings and vacua. The basic trick is to exploit the coset structure of the manifold describing the scalar fields to work always at the `origin' in field space, where it is easier to solve the quadratic constraints on the embedding tensor and to identify consistency conditions, gauge group, vacuum energy density and mass spectrum. So doing, many new gaugings and vacua were identified in \cite{di}, including some that lead to $N=0$ Minkowski vacua and were further studied in \cite{dz}. 

The features of this wider class of vacua include: possible tachyonic instabilities when moving along the multiple flat directions of the classical potential; at least four massless vectors and six massless scalars in the `physical' spectrum, i.e. after removing the would-be Goldstone bosons of the spontaneously broken gauge group; Dirac-type masses for the spin-$3/2$ and the spin-$1/2$ fermions in the spectrum. It is not obvious that these features will persist once the examples of \cite{di} are further generalised, but in the known examples an intriguing pattern is emerging. In particular, the full mass spectrum depends only on the charges of the eight supersymmetry generators with respect to at least one and at most four unbroken U(1) factors in the gauge group. Denoting by $\vec{q}_i \equiv (q_i^1,\ldots,q_i^n)$ the vector of the charges of $Q_i$ with respect to U(1)$^n$, in the models studied so far the 8 supercharges always come in pairs, with opposite charge vectors ($\vec{q}_1 = - \vec{q}_2$, $\vec{q}_3 = - \vec{q}_4$, $\vec{q}_5 = - \vec{q}_6$, $\vec{q}_7 = - \vec{q}_8$). This uniquely determines the charges of all the states, according to the following scheme:
\begin{eqnarray}
| +2 \rangle : & & \vec{0} \, , 
\nonumber \\
| +3/2 \, , \, i \rangle =  Q_i \, | +2 \rangle : & & \vec{q}_i \, , 
\nonumber \\
| +1 \, , \, [ij] \rangle =  Q_i \, Q_j | +2 \rangle : & & \vec{q}_i + \vec{q}_j \, , 
\nonumber \\
| +1/2 \, , \, [ijk] \rangle =  Q_i \, Q_j \, Q_k | +2 \rangle : & & \vec{q}_i + \vec{q}_j + \vec{q}_k \, , 
\nonumber \\
| 0 \, , \, [ijkl] \rangle =  Q_i \, Q_j \, Q_k \, Q_l | +2 \rangle : & & \vec{q}_i + \vec{q}_j + \vec{q}_k + \vec{q}_l \, , 
\label{U1charges}
\end{eqnarray}
and similarly for the CPT-conjugates. Acting with all 8 supercharges on the graviton of helicity $+2$ gives back the graviton of helicity $-2$, which leads to the constraint $\sum_{i=1}^8 \vec{q}_i = \vec{0}$. The spectrum is then given by:
\begin{eqnarray}
|2 \rangle : & & M^2 = 0 \, , 
\nonumber \\
|3/2 \, , \, i \rangle : & & M_i^2 = \left( \vec{q}_i \right)^2 \, , 
\nonumber \\
|1 \, , \, [ij] \rangle : & & M_{ij}^2 = \left( \vec{q}_i + \vec{q}_j \right)^2  \, , 
\nonumber \\
|1/2 \, , \, [ijk] \rangle : & & M_{ijk}^2 = \left( \vec{q}_i + \vec{q}_j + \vec{q}_k \right)^2  \, , 
\nonumber \\
| 0 \, , \, [ijkl] \rangle : & & M_{ijkl}^2 = \left( \vec{q}_i + \vec{q}_j + \vec{q}_k + \vec{q}_l \right)^2 \, , 
\label{U1spectrum}
\end{eqnarray}
where the scalar products of the charge vectors must be taken with a suitable field-dependent real diagonal metric, not necessarily positive definite (having it positive definite, however, guarantees the absence of tachyons for the corresponding field configurations).

On the basis of the above results, we showed in \cite{dz}  that for all these gaugings not only
\begin{equation}
{\rm Str} \, {\cal M}^6 = 0 \, , 
\label{result2}
\end{equation}
but also
\begin{equation}
{\rm Str} \, {\cal M}^8 > 0 \, . 
\label{result3}
\end{equation}
We also observed that, in contrast with the positive semi-definite classical potential generated by the CSS gauging, the potentials associated with the new gaugings of \cite{di}  can have tachyonic instabilities along their flat directions, and we identified those flat directions that do not lead to tachyons in the classical spectrum. After specializing to this part of the classical moduli space, we studied the one-loop effective potential, and we found analytical and numerical evidence that for all these field configurations, as a consequence of Eqs.~(\ref{result1}), (\ref{result2}) and (\ref{result3}), it is negative definite: 
\begin{equation}
V_1 < 0 \, . 
\label{result4}
\end{equation}
This means that at the one-loop level no locally stable vacua are known with fully broken supersymmetry and positive or vanishing vacuum energy. We stress that all the results mentioned above are highly non-trivial extensions of the known results valid for the CSS gauging: for general gaugings, and in particular for all the other explicit gaugings considered in \cite{dz}, we cannot rely on the group-theoretical explanation that was shown to be at the root of the supertrace formulas valid for the CSS gauging. 	

\section{Conclusions and outlook}

In conclusion, we proved in \cite{dz} that the one-loop effective potential of spontaneously broken
$N=8$ supergravity is calculable and finite at {\em all} classical
four-dimensional Minkowski vacua without tachyons in the spectrum, 
since the supertraces of the quadratic and quartic mass
matrices vanish along the classically flat directions: ${\rm Str} \,
{\cal M}^2 = {\rm Str} \, {\cal M}^4 =0$. We also showed that ${\rm Str}
\, {\cal M}^6 = 0$ but ${\rm Str} \, {\cal M}^8 > 0$ in a broad class
of vacua with broken supersymmetry on a flat background, which
includes all those explicitly identified so far, and  we found analytical
and numerical evidence that the corresponding one-loop effective
potential is negative definite. As a result, no locally stable 1-loop 
Minkowski or de~Sitter vacuum of fully broken theory is known to date.

It would be interesting to further enlarge the class of gaugings
leading to Minkowski (or de~Sitter) vacua, for example finding some
where the spectrum is not controlled by the charges with respect to
the unbroken $U(1)$ factors in the gauge group, and to check whether
the properties ${\rm Str} \, {\cal M}^6 = 0$ and ${\rm Str} \, {\cal
  M}^8 > 0$ still hold. It would be also interesting to establish more
firmly the connection between  ${\rm Str} \, {\cal
  M}^8 > 0$ and $V_1 < 0$. Finally, it would be important to
understand the nature of the obstruction that has so far prevented the
discovery of locally stable dS vacua of the theory, with the goal of
either finding an example or proving a general no-go theorem.

Other groups are also working on the subject and further progress may
come soon.

\end{document}